# Some differences between Dirac's hole theory and quantum field theory

by

Dan Solomon

Rauland-Borg Corporation
3450 W Oakton
Skokie, IL 60076
USA

Email: dan.solomon@rauland.com

July 22, 2004

**Abstract**

Dirac's hole theory (HT) and quantum field theory (QFT) are generally considered to be equivalent to each other.  However, it has been recently shown by several researchers that this is not necessarily the case.  When the change in the vacuum energy was calculated for a time independent perturbation HT and QFT yielded different results.  In this paper we extend this discussion to include a time dependent perturbation for which the exact solution to the Dirac equation is known.  It will be shown that for this case, also, HT and QFT yield different results.  In addition, there will be some discussion of the problem of anomalies in QFT.

# I. Introduction.

Dirac's hole theory and quantum field theory are generally assumed to be equivalent. Recently several papers have appeared in the literature pointing out that there are differences between Dirac's hole theory (HT) and quantum field theory (QFT) [1][2][3][4][5]. The problem was originally examined by Coutinho et al[1][2]. They calculated the second order change in the energy of the vacuum state due to a time independent perturbation. They found that HT and QFT produce different results. They concluded that the difference between HT and QFT was related to the validity of Feynman's belief that the Pauli Exclusion Principle can be disregarded for intermediate states in perturbation theory. This belief was based on Feynman's observation that terms that violate the Pauli principle formally cancel out in perturbation theory. However Coutino et al show that this is not necessarily the case for HT when applied to an actual problem. This author (Solomon [4]) found that this problem was related to the way the vacuum state was defined in QFT. If the definition of the vacuum state was modified as described in [4] then the HT and QFT would yield identical results.

Most of the previous work [1-4] examined the difference between HT and QFT for a time independent change in the electric potential. In this article the differences between HT and QFT for a time dependent change in the electric potential will be examined. Perturbation theory is normally used to work this type of problem because it is generally not possible to find exact solutions to the Dirac equation for time dependent changes in the electric potential. However, in this article we will examine the change in the energy of a HT quantum state versus a QFT quantum state due to a time dependent electric potential for which the exact solution to the Dirac equation is known. It will be shown that the HT solution is different from the QFT solution. We will also consider the implications for QFT due the differences with HT. It will be shown that these differences are related to the problem of anomalies in QFT.

In order to simplify the discussion and avoid unnecessary mathematical details we will work in 1-1 dimensional space-time where the space dimension is taken along the z-axis and use natural units so that $\hbar = c = 1$. In this case the Dirac equation for a single electron is,

$$i\frac{\partial \psi(z,t)}{\partial t} = H\psi(z,t) \tag{1}$$

where the Dirac Hamiltonian is given by,

$$H = H_0 + qV \tag{2}$$

where $H_0$ is the Hamiltonian in the absence of interactions and V is an external perturbation. For the 1-1D case,

$$\hat{H}_0 = \left(-i\sigma_x \frac{\partial}{\partial z} + m\sigma_z\right) \text{ and } V = -\sigma_x A_z + A_0 \tag{3}$$

where $\sigma_x$ and $\sigma_z$ are the usual Pauli matrices and $(A_0, A_z)$ is the electric potential which, for the purposes of this discussion, are assumed to be classical, real valued quantities. We will assume periodic boundary conditions so that the solutions satisfy $\psi(z,t) = \psi(z+L,t)$ where L is the 1-dimensional integration volume. In this case the orthonormal free field solutions (V is zero) of (1) are given by,

$$\varphi_{\lambda,r}^{(0)}(z,t) = \varphi_{\lambda,r}^{(0)}(z)e^{-i\varepsilon_{\lambda,r}^{(0)}t} = u_{\lambda,r}e^{-i\left(\varepsilon_{\lambda,r}^{(0)}t - p_r z\right)} \tag{4}$$

where 'r' is an integer, $\lambda = \pm 1$ is the sign of the energy, $p_r = 2\pi r/L$, and where,

$$\varepsilon_{\lambda,r}^{(0)} = \lambda E_r \,; \; E_r = +\sqrt{p_r^2 + m^2} \,; \; u_{\lambda,r} = N_{\lambda,r} \begin{pmatrix} 1 \\ p_r / (\lambda E_r + m) \end{pmatrix} \,; \; N_{\lambda,r} = \sqrt{\frac{\lambda E_r + m}{2L\lambda E_r}} \quad (5)$$

The quantities $\varphi_{\lambda,r}^{(0)}(z)$ satisfy the relationship,

$$\hat{H}_0 \varphi_{\lambda,r}^{(0)}(z) = \varepsilon_{\lambda,r}^{(0)} \varphi_{\lambda,r}^{(0)}(z) \quad (6)$$

The $\varphi_{\lambda,r}^{(0)}(z)$ form an orthonormal basis set and satisfy,

$$\int_{-L/2}^{+L/2} \varphi_{\lambda,r}^{(0)\dagger}(z) \varphi_{\lambda',s}^{(0)}(z) dz = \delta_{\lambda\lambda'} \delta_{rs} \quad (7)$$

The energy $\xi(\psi(z,t))$ of a normalized wave function $\psi(z,t)$ is given by,

$$\xi(\psi(z,t)) = \int_{-L/2}^{+L/2} \psi^\dagger(z,t)(H_0 + qV)\psi(z,t) dz \quad (8)$$

The free field energy $\xi_f(\psi(z,t))$ for the normalized wave function $\psi(z,t)$ is defined by,

$$\xi_f(\psi(z,t)) = \int_{-L/2}^{+L/2} \psi^\dagger(z,t) H_0 \psi(z,t) dz \quad (9)$$

This is similar to the energy but with the interaction term V left out. It is evident that when $V = 0$ the free field energy and the "total" energy are equivalent. Therefore, in the following discussion if we know that $V = 0$ we will use the symbol $\xi_f$, instead of $\xi$, for the energy.

## **II. Hole Theory**

The fact that there are negative energy solutions to the Dirac equation creates a problem because positive energy electrons would not be stable. They would tend to decay into negative energy states. To solve this dilemma Dirac proposed that each negative energy state was occupied by a single electron. In this case the Pauli exclusion principle would prevent a positive energy electron from decaying into a negative energy state. Therefore Dirac theory is an N-electron theory where $N \to \infty$. In this paper we will assume that these electrons are non-interacting.

For an N-electron theory the wave function is written as a Slater determinant [6,7,8],

$$\Psi^N(z_1, z_2, ..., z_N, t) = \frac{1}{\sqrt{N!}} \sum_P (-1)^p P\left(\psi_1(z_1, t) \psi_2(z_2, t) \cdots \psi_N(z_N, t)\right) \quad (10)$$

where the $\psi_n(z, t)$ ($n = 1, 2, ..., N$) are a normalized and orthogonal set of wave functions that obey the Dirac equation, P is a permutation operator acting on the space coordinates, and p is the number of interchanges in P. Note if $\psi_a(z, t)$ and $\psi_b(z, t)$ are two wave functions that obey the Dirac equation then it can be shown that,

$$\frac{\partial}{\partial t} \int_{-L/2}^{+L/2} \psi_a^\dagger(z, t) \psi_b(z, t) dz = 0 \quad (11)$$

Therefore if the $\psi_n(z, t)$ in (10) are orthogonal at some initial time then they are orthogonal for all time.

The expectation value of a single particle operator $O_{op}(z)$ is defined as,

$$O_e = \int \psi^\dagger(z, t) O_{op}(z) \psi(z, t) dz \quad (12)$$

where $\psi(z, t)$ is a normalized single particle wave function. The N-electron operator is given by,

$$O_{op}^N(z_1, z_2, ..., z_N) = \sum_{n=1}^{N} O_{op}(z_n) \quad (13)$$

which is just the sum of one particle operators. The expectation value of a normalized N-electron wave function is,

$$O_e^N = \int \Psi^{N\dagger}(z_1, z_2, ..., x_N, t) O_{op}^N(z_1, z_2, ..., z_N) \Psi^N(z_1, z_2, ..., z_N, t) dz_1 dz_2 ... dz_N \quad (14)$$

This can be shown to be equal to,

$$O_e^N = \sum_{n=1}^{N} \int \psi_n^\dagger(z, t) O_{op}(z) \psi_n(z, t) d\vec{x} \quad (15)$$

That is, the N electron expectation value is just the sum of the single particle expectation values associated with each of the individual wave functions $\psi_n$. For example, the free field energy $\xi_f(\Psi^N)$ of the N-electron state is,

$$\xi_f(\Psi^N) = \sum_{n=1}^{N} \int \psi_n^\dagger(z, t) H_0 \psi_n(z, t) dz = \sum_{n=1}^{N} \xi_f(\psi_n) \quad (16)$$

Assume, at some initial time $t_0$, the electric potential is zero and the system is in some initial unperturbed state. In HT the unperturbed vacuum state is the state where each negative energy wave function $\varphi_{-1,r}^{(0)}$ is occupied by a single electron and each positive energy state $\varphi_{+1,r}^{(0)}$ is unoccupied. The energy of the vacuum state is given by summing over the energies of all the negative energy states. The total energy of the unperturbed vacuum state is then,

$$E_{hvac}^{(0)} = \sum_r \varepsilon_{-1,r}^{(0)} = -\sum_r E_r \quad (17)$$

We can add an additional electron provided it consists of a combination of positive energy states $\varphi_{+1,r}^{(0)}$ so that it is orthogonal to the vacuum wave functions $\varphi_{-1,r}^{(0)}$. Let the wave function that defines this positive energy electron, at the initial time $t_0$, be given by,

$$\psi_p(z,t_0) = \sum_r f_r \varphi_{+1,r}^{(0)}(z,t_0) \tag{18}$$

where the $f_r$ are constant expansion coefficients. Assume that the $f_r$ are selected so that $\psi_p(z,t_0)$ is normalized.

Define then, at the initial time $t_0$, the initial state $S(t_0)$ consisting of the unperturbed vacuum electrons $\varphi_{-1,r}^{(0)}(z,t_0)$ and a single positive energy electron $\psi_p(z,t_0)$. Therefore the energy of $S(t_0)$ is,

$$E_T(t_0) = \xi_f(\psi_p(z,t_0)) + E_{hvac}^{(0)} \tag{19}$$

where we have used the fact that the electric potential is zero.

Now we are not really interested in the total energy but in the energy with respect to the unperturbed vacuum state. Therefore we subtract the vacuum energy $E_{hvac}^{(0)}$ from the above expression to obtain,

$$E_{T,R}(t_0) = E_T(t_0) - E_{hvav}^{(0)} = \xi_f(\psi_p(z,t_0)) \tag{20}$$

which is just the energy of the positive energy electron.

Next, consider the change in the energy due to an interaction with an external electric potential. At the initial time $t_0$ the electric potential is zero and the system is in the unperturbed initial state. Next apply an electric potential and then remove it at some later time $t_1$ so that,

$$(A_0, A_z) = 0 \text{ for } t < t_0; \ (A_0, A_z) \neq 0 \text{ for } t_0 \leq t \leq t_1; \ (A_0, A_z) = 0 \text{ for } t > t_1 \qquad (21)$$

Now what is the change in the energy of the quantum system S due to this interaction with the electric potential? Under the action of the electric potential each wave function $\varphi_{\lambda,r}^{(0)}(z, t_0)$ evolves into the final state $\varphi_{\lambda,r}(z, t_f)$ where $t_f > t_1$. Also the the wave function $\psi_p(z, t_0)$ evolves into $\psi_p(z, t_f)$. Note that per equation (21) the electric potential is zero at the final time $t_f > t_1$. Therefore the change in the energy of each vacuum electron is,

$$\Delta\varepsilon_{-1,r} = \int_{-L/2}^{+L/2} \varphi_{-1,r}^\dagger(z, t_f) H_0 \varphi_{-1,r}(z, t_f) dz - \varepsilon_{-1,r}^{(0)} \qquad (22)$$

The change in the energy of the positive energy electron is,

$$\Delta\xi_{fp} = \xi_f(\psi_p(z, t_f)) - \xi_f(\psi_p(z, t_0)) \qquad (23)$$

The total change in the energy of the system S is then,

$$\Delta E_T = \Delta\xi_{fp} + \sum_r \Delta\varepsilon_{-1,r} \qquad (24)$$

Therefore the energy of the system S at $t_f$ with respect to the unperturbed vacuum state is,

$$E_{T,R}(t_f) = E_{T,R}(t_0) + \Delta E_T = \xi_f(\psi_p(z, t_0)) + \Delta E_T = \xi_p(\psi_p(z, t_f)) + \Delta E_{hvac} \qquad (25)$$

where $\Delta E_{hvac}$ is the change in the energy associated with the vacuum electrons and is given by,

$$\Delta E_{hvac} = \sum_r \Delta\varepsilon_{-1,r} \qquad (26)$$

From the above discussion the electric potential is zero for $t < t_0$ and $t > t_1$. For $t_0 \leq t \leq t_1$ let the electric potential be given by,

$$(A_0, A_z) = \left(\frac{\partial \chi}{\partial t}, -\frac{\partial \chi}{\partial z}\right) \text{ for } t_0 \leq t \leq t_1 \tag{27}$$

where $\chi(z,t)$ is an arbitrary real valued function that satisfies the initial conditions at $t = t_0$, $\frac{\partial \chi(z,t_0)}{\partial t} = 0$ and $\chi(z,t_0) = 0$. Now given an initial wave function $\psi(z,t_0)$ at time $t_0$ what is the final wave function $\psi(z,t_f)$ at some final time $t_f > t_1$. Use (27), (21), and (2) in (1) to obtain,

$$i\frac{\partial \psi}{\partial t} = \left(H_0 + q\sigma_x \frac{\partial \chi}{\partial z} + q\frac{\partial \chi}{\partial t}\right)\psi \text{ for } t_0 \leq t \leq t_1 \tag{28}$$

and,

$$i\frac{\partial \psi}{\partial t} = H_0 \psi \text{ for } t > t_1 \tag{29}$$

Since the time derivative is to the first order the boundary condition at $t = t_1$ is,

$$\psi(z, t_1 + \delta) \underset{\delta \to 0}{=} \psi(z, t_1 - \delta) \tag{30}$$

The solution to (28) is,

$$\psi(z,t) = e^{-iq\chi(z,t)} e^{-iH_0(t-t_0)} \psi(z,t_0) \text{ for } t_0 \leq t \leq t_1 \tag{31}$$

The solution to (29) is,

$$\psi(z,t_f) = e^{-iH_0(t_f-t_1)} \psi(z,t_1) \text{ for } t_f > t_1 \tag{32}$$

Using the boundary conditions (30) we obtain,

$$\psi(z,t_f) = e^{-iH_0(t_f-t_1)} e^{-iq\chi(z,t_1)} e^{-iH_0(t_1-t_0)} \psi(z,t_0) \tag{33}$$

The energy of the final state $\psi(z, t_f)$ is,

$$\xi_f\left(\psi(z, t_f)\right) = \int_{-L/2}^{+L/2} \psi^\dagger(z, t_0) e^{+iH_0(t_1-t_0)} \left(H_0 - q\sigma_x \frac{\partial \chi(z, t_1)}{\partial z}\right) e^{-iH_0(t_1-t_0)} \psi(z, t_0) dz \tag{34}$$

where we have used $e^{+iq\chi} H_0 e^{-iq\chi} = \left(H_0 - q\sigma_x \partial\chi/\partial z\right)$. This becomes,

$$\xi_f\left(\psi(z, t_f)\right) = \xi_f\left(\psi(z, t_0)\right) - q \int_{-L/2}^{+L/2} \frac{\partial \chi(z, t_1)}{\partial z} \left(\psi^{(0)\dagger}(z, t_1) \sigma_x \psi^{(0)}(z, t_1)\right) dz \tag{35}$$

where,

$$\psi^{(0)}(z, t_1) = e^{-iH_0(t_1-t_0)} \psi(z, t_0) \tag{36}$$

Integrate by parts and assume reasonable boundary conditions to obtain,

$$\xi_f\left(\psi(z, t_f)\right) = \xi_f\left(\psi(z, t_0)\right) + q \int_{-L/2}^{+L/2} \chi(z, t_1) \frac{\partial}{\partial z} \left(\psi^{(0)\dagger}(z, t_1) \sigma_x \psi^{(0)}(z, t_1)\right) dz \tag{37}$$

Therefore the change in the energy from $t_0$ to $t_f$ is,

$$\Delta \xi_f = q \int_{-L/2}^{+L/2} \chi(z, t_1) \frac{\partial}{\partial z} \left(\psi^{(0)\dagger}(z, t_1) \sigma_x \psi^{(0)}(z, t_1)\right) dz \tag{38}$$

Based on the above discussion each negative energy state $\varphi_{-1,r}^{(0)}(z, t_0)$ evolves into the state,

$$\varphi_{-1,r}(z, t_f) = e^{-iH_0(t_f-t_1)} e^{-iq\chi(z,t_1)} e^{-iH_0(t_1-t_0)} \varphi_{-1,r}^{(0)}(z, t_0) \tag{39}$$

and the change in the energy of this state from $t_0$ to $t_f$ is

$$\Delta \varepsilon_{-1,r} = q \int_{-L/2}^{+L/2} \chi(z, t_1) \frac{\partial}{\partial z} \left(\varphi_{-1,r}^{(0)\dagger}(z, t_1) \sigma_x \varphi_{-1,r}^{(0)}(z, t_1)\right) dz \tag{40}$$

From (4) we can show that,

$$\varphi^{(0)\dagger}_{-1,r}(z,t_1)\sigma_x\varphi^{(0)}_{-1,r}(z,t_1) = u^{\dagger}_{-1,r}\sigma_x u_{-1,r} \tag{41}$$

The above expression is independent of z. Therefore the derivative with respect to z in (40) is zero so that,

$$\Delta\varepsilon_{-1,r} = 0 \tag{42}$$

From the discussion leading up to (34) the wave function $\psi_p$ at the final time $t_f$ is,

$$\psi_p(z,t_f) = e^{-iH_0(t_f-t_1)}e^{-iq\chi(z,t_1)}e^{-iH_0(t_1-t_0)}\psi_p(z,t_0) \tag{43}$$

and the difference in energy between the final state $\psi_p(z,t_f)$ and initial state $\psi_p(z,t_0)$ is,

$$\Delta\xi_{fp} = \xi_f\left(\psi_p(z,t_f)\right) - \xi_f\left(\psi_p(z,t_0)\right) = \int_{-L/2}^{+L/2} \chi(z,t_1)\frac{\partial J^{(0)}_p(z,t_1)}{\partial z}dz \tag{44}$$

where,

$$J^{(0)}_p(z,t_1) = q\left(\psi^{(0)\dagger}_p(z,t_1)\sigma_x\psi^{(0)}_p(z,t_1)\right) \tag{45}$$

with,

$$\psi^{(0)}_p(z,t_1) = e^{-iH_0(t_1-t_0)}\psi_p(z,t_0) \tag{46}$$

The quantity $\psi^{(0)}_p(z,t_1)$ is the wavefunction that $\psi_p(z,t_0)$ would evolve into by the time $t_1$ if the electric potential was zero. Note that $J^{(0)}_p(z,t_1)$ is independent from $\chi(z,t)$. Therefore variations in $\chi(z,t)$ do not effect $J^{(0)}_p(z,t_1)$. Now suppose that the initial wave function $\psi_p(z,t_0)$ is selected so that $\dfrac{\partial J^{(0)}_p(z,t_1)}{\partial z} \neq 0$. An example of this is given in the Appendix. If this condition is satisfied then we can find a $\chi(z,t_1)$ such that $\xi_f(\psi_p(z,t_f))$ is negative number with an arbitrarily large magnitude. For instance let

$$\chi(z,t_1) = -\lambda \frac{\partial J_p^{(0)}(z,t_1)}{\partial z}$$ where $\lambda$ is a constant. Use this in (44) to obtain,

$$\xi_f(\psi_p(z,t_f)) = \xi_f(\psi_p(z,t_0)) - \lambda \int_{-L/2}^{+L/2} \left(\frac{\partial J_p^{(0)}(z,t_1)}{\partial z}\right)^2 dz \qquad (47)$$

Therefore as $\lambda \to \infty$, $\xi_f(\psi_p(z,t_f)) \to -\infty$. Use (42) in (26) to obtain $\Delta E_{hvac} = 0$. Use this result along with (47) in (25) to obtain,

$$E_{T,R}(t_f) = \xi_f(\psi_p(z,t_0)) - \lambda \int_{-L/2}^{+L/2} \left(\frac{\partial J_p^{(0)}(z,t_1)}{\partial z}\right)^2 dz \qquad (48)$$

$E_{T,R}(t_f)$ is the energy with respect to the unperturbed vacuum at the final time $t_f$. As $\lambda \to \infty$, $E_{T,R}(t_f) \to -\infty$. Therefore the energy of the quantum system S at the time $t_f$ can be less than that of the unperturbed vacuum state by an arbitrarily large amount.

### III. Discussion

This result is somewhat surprising. It shows that in HT the unperturbed vacuum state is not the lowest energy state and that it is possible to extract an unlimited amount of energy from an initial quantum state. To review the results of the previous section we started with an initial system consisting of vacuum electrons in their unperturbed state $\varphi_{-1,r}^{(0)}$ and a positive energy electron $\psi_p$ as defined by (18). We apply the electric potential described by equations (21) and (27). The result is that each wave function evolves from its initial state in accordance with (33). We find that the change in energy of the vacuum electrons from the initial to final state is zero. However the change in energy of the wave function $\psi_p$ can be a negative number with an arbitrarily large magnitude. The net result is that the total energy of the system is negative with respect to the unperturbed vacuum energy.

In the above example the energy of the vacuum electrons doesn't change and the energy of the wave function $\psi_p$, which was originally positive, becomes negative. Now wasn't the Pauli principle suppose to prevent this? The Pauli principle is simply the statement that no more than one electron can occupy a given state at given time. Equations (10) and (11) are the mathematical realization of this principle. The Pauli Principle is a result of the fact that if the initial wave functions are orthogonal then the wave functions will be orthogonal for all time. Therefore two electrons cannot end up in the same state. In the problem discussed in Section II the change in the wave function is given by (see Eq. (33)),

$$\psi(z, t_f) = U\psi(z, t_0) \tag{49}$$

where,

$$U = e^{-iH_0(t_f - t_1)} e^{-iq\chi(z, t_1)} e^{-iH_0(t_1 - t_0)} \tag{50}$$

Therefore if,

$$\psi_a(z, t_f) = U\psi_a(z, t_0) \text{ and } \psi_b(z, t_f) = U\psi_b(z, t_0) \tag{51}$$

then,

$$\int_{-L/2}^{+L/2} \psi_a^\dagger(z, t_f) \psi_b(z, t_f) dz = \int_{-L/2}^{+L/2} \psi_a^\dagger(z, t_0) U^\dagger U \psi_b(z, t_0) dz = \int_{-L/2}^{+L/2} \psi_a^\dagger(z, t_0) \psi_b(z, t_0)$$

$$\tag{52}$$

Thus the transformation (49) preserves the orthogonality of the wave functions and is consistent with the Pauli principle. Therefore the conjecture that the Pauli principle eliminates the possibility of quantum states existing with less energy than that of the unperturbed vacuum state is not correct.

### IV. Quantum field theory.

In the Section II we derived an expression for the change in the free field energy in HT. It was shown that the final energy is less than the energy of the unperturbed vacuum state. Now we want to work the same problem using quantum field theory. We shall work in the Schrödinger picture. In this case the field operators are time independent and all changes in the system are reflected in the changes of the state vectors. The field operators are defined by,

$$\hat{\psi}(z) = \sum_r \left( \hat{b}_r \varphi^{(0)}_{+1,r}(z) + \hat{d}^\dagger_r \varphi^{(0)}_{-1,r}(z) \right); \quad \hat{\psi}^\dagger(z) = \sum_r \left( \hat{b}^\dagger_r \varphi^{(0)\dagger}_{1,r}(z) + \hat{d}_r \varphi^{(0)\dagger}_{-1,r}(z) \right) \qquad (53)$$

where the $\hat{b}_r$ and $\hat{b}^\dagger_r$ are the electron destruction and creation operators, respectively associated with the state $\varphi^{(0)}_{+1,r}$ and the $\hat{d}_r$ and $\hat{d}^\dagger_r$ are the positron destruction and creation operators, respectively, associated with the state $\varphi^{(0)}_{-1,r}$. The destruction and creation operators obey the following relationships,

$$\hat{b}_m \hat{b}^\dagger_n + \hat{b}^\dagger_n \hat{b}_m = \hat{d}_m \hat{d}^\dagger_n + \hat{d}^\dagger_n \hat{d}_m = \delta_{mn}; \quad \text{all other anticommutators=0} \qquad (54)$$

The Hamiltonian operator is,

$$\hat{H} = \hat{H}_0 + q\hat{V} \qquad (55)$$

where,

$$\hat{H}_0 = \int_{-L/2}^{+L/2} \hat{\psi}^\dagger H_0 \hat{\psi} dx - \xi_{ren} \text{ and } \hat{V} = \int_{-L/2}^{+L/2} \hat{\psi}^\dagger V \hat{\psi} dx \tag{56}$$

where $\xi_{ren}$ is a renormalization constant defined so that the energy of the vacuum state $|0\rangle$ is equal to zero.

The time dependent state vector $|\Omega(t)\rangle$ and its dual $\langle\Omega(t)|$ obey,

$$i\frac{\partial|\Omega(t)\rangle}{\partial t} = \hat{H}|\Omega(t)\rangle; \quad -i\frac{\partial\langle\Omega(t)|}{\partial t} = \langle\Omega(t)|\hat{H} \tag{57}$$

From the above discussion we obtain,

$$\hat{H}_0 = \sum_r E_r \left(\hat{b}_r^\dagger \hat{b}_r - \hat{d}_r \hat{d}_r^\dagger\right) - \xi_{ren} = \sum_r E_r \left(\hat{b}_r^\dagger \hat{b}_r + \hat{d}_r^\dagger \hat{d}_r\right) \tag{58}$$

where the last step is obtained by using (54) and by properly defining $\xi_{ren}$.

The vacuum state $|0\rangle$ is the quantum state which is destroyed by the positron and electron destruction operators, i.e.,

$$\hat{d}_n|0\rangle = \hat{b}_n|0\rangle = 0 \tag{59}$$

The vacuum state satisfies the equation,

$$\hat{H}_0|0\rangle = 0 \tag{60}$$

Therefore $|0\rangle$ is an eigenstate of the operator $\hat{H}_0$ with an eigenvalue $\varepsilon(|0\rangle) = 0$. Additional eigenstates $|k_j\rangle$ are produced by acting on $|0\rangle$ with the electron and positron creation operators $\hat{b}_n^\dagger$ and $\hat{d}_n^\dagger$. The effect of the action of these creation operators is to increase the energy of the initial state. Therefore eigenstates $|k_j\rangle$ satisfy,

$$\hat{H}_0|k_j\rangle = \varepsilon(|k_j\rangle)|k_j\rangle \text{ where } \varepsilon(|k_j\rangle) > \varepsilon(|0\rangle) = 0 \text{ if } |k_j\rangle \neq |0\rangle \tag{61}$$

These eigenstates $|k_j\rangle$ form an orthonormal set so that,

$$\langle k_j | k_i \rangle = \delta_{ij} \text{ and } \sum_j |k_j\rangle\langle k_j| = 1 \tag{62}$$

Any arbitrary state $|\Omega\rangle$ can be expressed as a Fourier sum of the eigenstates $|\Omega\rangle$,

$$|\Omega\rangle = \sum_j c_j |k_j\rangle \tag{63}$$

If the electric potential is zero then the energy of a normalized state vector $|\Omega\rangle$ is,

$$E(|\Omega\rangle) = \langle \Omega | \hat{H}_0 | \Omega \rangle \tag{64}$$

Using the above relationships it is easy to show that,

$$E(|\Omega\rangle) \geq \varepsilon(|0\rangle) = 0 \text{ for all } |\Omega\rangle \tag{65}$$

Now suppose that at the initial time $t_0$ the state vector is $|\Omega(t_0)\rangle$. Next apply an electric potential per (21). At time $t_f > t_1$ the state vector is $|\Omega(t_f)\rangle$. Now from (65) the energy of $|\Omega(t_f)\rangle$ must be greater than or equal to the energy of the unperturbed vacuum state $|0\rangle$. In QFT it is not possible to interact with an electric potential to produce a quantum state with less energy than the vacuum state because there does not exist, within the theory, a state vector $|\Omega\rangle$ whose free field energy is less than that of the vacuum state $|0\rangle$. This is in sharp contrast to HT where, as we have shown, it is possible to produce a quantum system with less energy than the unperturbed vacuum state. Therefore HT and QFT produce different results.

The fact that the vacuum state is the lowest energy state appears to be a desirable feature and would seem to make QFT superior to HT. However, as will be shown, this feature comes with a price. As was pointed out in Section II HT is an N-electron theory where each electron wave function evolves independently in time according to the Dirac equation. Any observable quantity, such as current, charge, or energy, is the sum of the corresponding observable for each wave function. Therefore all the symmetries and conservation laws associated with the Dirac equation hold for HT. Since HT and QFT are not equivalent the obvious question that arises is do these symmetries and conservation laws hold for QFT? This question will be addressed in the next section.

### **V. Anomalies in QFT.**

Another important area in which HT and QFT differ is in the area of anomalies. An anomaly occurs when the result of some calculation is not consistent with some symmetry of the Dirac equation. Consider, for example, the continuity equation. For the single particle wave function $\psi$ the charge and current are defined by,

$$\rho = q\psi^\dagger \psi \text{ and } J = q\psi^\dagger \sigma_x \psi \tag{66}$$

Using the Dirac equation it is easy to show that,

$$\frac{\partial \rho}{\partial t} + \frac{\partial J}{\partial z} = 0 \tag{67}$$

This is called the continuity equation. For an N-electron theory the total current and charge are,

$$\rho_N = q \sum_{n=1}^{N} \psi_n^\dagger \psi_n \text{ and } J_N = q \sum_{n=1}^{N} \psi_n^\dagger \sigma_x \psi_n \tag{68}$$

Each of the wave functions $\psi_n$ obeys the Dirac equation and therefore satisfies the continuity equation. From this it is evident that,

$$\frac{\partial \rho_N}{\partial t} + \frac{\partial J_N}{\partial z} = 0 \tag{69}$$

Thus an N-electron theory obeys the continuity equation. Therefore the continuity equation holds in HT.

Now we will examine the situation in QFT in the Schrödinger picture. Here we define the charge and current operators as,

$$\hat{\rho} = q\hat{\psi}^{\dagger}\hat{\psi} \text{ and } \hat{J} = q\hat{\psi}^{\dagger}\sigma_x\hat{\psi} \tag{70}$$

Use this in (56) and (55) along with (3) to obtain,

$$\hat{H} = \hat{H}_0 - \int_{-L/2}^{+L/2} \hat{J}(z) A_z(z,t) dz + \int_{-L/2}^{+L/2} \hat{\rho}(z) A_0(z,t) dz \tag{71}$$

For a normalized state vector $|\Omega\rangle$ the current and charge expectation values are,

$$\rho_e = \langle \Omega|\hat{\rho}|\Omega\rangle \text{ and } J_e = \langle \Omega|\hat{J}|\Omega\rangle \tag{72}$$

In QFT the continuity equation is given by,

$$\frac{\partial \rho_e}{\partial t} + \frac{\partial J_e}{\partial z} = 0 \tag{73}$$

We want to determine if the above relationship is true. To determine this start by using (57) and (72) to obtain,

$$\frac{\partial \rho_e}{\partial t} = \frac{\partial \langle \Omega|\hat{\rho}|\Omega\rangle}{\partial t} = i\langle \Omega|[\hat{H},\hat{\rho}]|\Omega\rangle \tag{74}$$

Use (71) in the above to yield,

$$\frac{\partial \rho_e(z,t)}{\partial t} = i \left\{ \begin{array}{l} \langle \Omega(t)|[\hat{H}_0,\hat{\rho}(z)]|\Omega(t)\rangle \\ -\int_{-L/2}^{+L/2} \langle \Omega(t)|[J(z'),\hat{\rho}(z)]|\Omega(t)\rangle A_z(z'.t)dz' \\ \int_{-L/2}^{+L/2} \langle \Omega(t)|[\hat{\rho}(z'),\hat{\rho}(z)]|\Omega(t)\rangle A_0(z',y)dz' \end{array} \right\} \tag{75}$$

Compare this relationship to (73). For (73) to be true for all possible values of $(A_0, A_z)$ and state vector $|\Omega\rangle$ the following relationships must hold,

$$i\left[\hat{H}_0, \hat{\rho}(z)\right] = -\frac{\partial J(z)}{\partial z} \tag{76}$$

$$\left[J(z'), \hat{\rho}(z)\right] = 0 \tag{77}$$

$$\left[\hat{\rho}(z'), \hat{\rho}(z)\right] = 0 \tag{78}$$

However it was shown by Schwinger [9] that (77) cannot be true. To show this take the derivative of the quantity $\left[\hat{J}(z'), \hat{\rho}(z)\right]$ with respect to $z'$ and use (76) to obtain,

$$\frac{\partial}{\partial z'}\left[\hat{J}(z'), \hat{\rho}(z)\right] = \left[\frac{\partial}{\partial z'}\hat{J}(z'), \hat{\rho}(z)\right] = -i\left[\left[\hat{H}_0, \hat{\rho}(z')\right], \hat{\rho}(z)\right] \tag{79}$$

Next expand the commutator to yield,

$$i\frac{\partial}{\partial z'}\left[\hat{J}(z'), \hat{\rho}(z)\right] = \hat{H}_0\hat{\rho}(z')\hat{\rho}(z) - \hat{\rho}(z')\hat{H}_0\hat{\rho}(z) - \hat{\rho}(z)\hat{H}_0\hat{\rho}(z') + \hat{\rho}(z)\hat{\rho}(z')\hat{H}_0 \tag{80}$$

Sandwich the above expression between the state vector $\langle 0|$ and its dual $|0\rangle$ and use $\hat{H}_0|0\rangle = 0$ and $\langle 0|\hat{H}_0 = 0$ to obtain,

$$-i\frac{\partial}{\partial z'}\langle 0|\left[\hat{J}(z'), \hat{\rho}(z)\right]|0\rangle = \langle 0|\hat{\rho}(z')\hat{H}_0\hat{\rho}(z)|0\rangle + \langle 0|\hat{\rho}(z)\hat{H}_0\hat{\rho}(z')|0\rangle \tag{81}$$

Next set $z = z'$ to obtain,

$$\frac{\partial}{\partial z'}\langle 0|\left[\hat{J}(z'), \hat{\rho}(z)\right]|0\rangle\bigg|_{z=z'} = 2i\langle 0|\hat{\rho}(z)\hat{H}_0\hat{\rho}(z)|0\rangle \tag{82}$$

Use (62) in the above to obtain,

$$\frac{\partial}{\partial z'}\langle 0|\left[\hat{J}(z'), \hat{\rho}(z)\right]|0\rangle\bigg|_{z=z'} = 2i\sum_{n,m}\langle 0|\hat{\rho}(z)|k_n\rangle\langle k_n|\hat{H}_0|k_m\rangle\langle k_m|\hat{\rho}(z)|0\rangle \tag{83}$$

Next use (61) to obtain,

$$\frac{\partial}{\partial z'}\langle 0|[\hat{J}(z'),\hat{\rho}(z)]|0\rangle\Big|_{z=z'} = 2i\sum_{n}\varepsilon(|k_n\rangle)\langle 0|\hat{\rho}(z)|k_n\rangle\langle k_n|\hat{\rho}(z)|0\rangle = 2i\sum_{n}\varepsilon(|k_n\rangle)|\langle 0|\hat{\rho}(z)|k_n\rangle|^2$$
(84)

Now, in general, the quantity $\langle 0|\hat{\rho}(z)|k_n\rangle$ is not zero [9] and since $\varepsilon(|k_n\rangle) > 0$ (except when $|k_n\rangle = |0\rangle$) the above expression is non-zero. Therefore the quantity $[\hat{J}(z'),\hat{\rho}(z)]$ cannot be zero and equation (77) is not valid. Therefore the continuity equation is not valid for QFT in the Schrödinger picture.

The result of this analysis is that we cannot assume that the symmetries of the Dirac equation hold for QFT so that we should expect that anomalies will occur in various calculations. And this is, indeed, the case. Consider the problem of gauge invariance. A change in the gauge is a change in the electric potential which does not produce a change the electromagnetic field (see discussion of this in [10] and [11]). A physical theory must be gauge invariant which means that all physical observables (such as the current or charge) are not affected by a gauge transformation. It is well known that when certain quantities are calculated in QFT, using standard perturbation theory, the results are not gauge invariant. The non-gauge invariant terms that appear in the results have to be removed to make the answer physically correct. A well known example of this is the calculation of the vacuum polarization tensor. Consider, for example, a calculation of the vacuum polarization tensor by Heitler (see page 322 of [12]). Heitler's solution for the Fourier transform of the vacuum polarization tensor is,

$$\pi^{uv}(k) = \pi^{uv}_G(k) + \pi^{uv}_{NG}(k)$$
(85)

The first term on the right hand side is given by,

$$\pi_G^{\mu\nu}(k) = \left(\frac{2q^2}{3\pi}\right)\left(k^\mu k^\nu - g^{\mu\nu}k^2\right)\int_{2m}^{\infty} dz \frac{\left(z^2+2m^2\right)\sqrt{\left(z^2-4m^2\right)}}{z^2\left(z^2-k^2\right)} \tag{86}$$

The second term on the right of (85) is

$$\pi_{NG}^{\mu\nu}(k) = \left(\frac{2q^2}{3\pi}\right)g_\nu^\mu\left(1-g^{\mu 0}\right)\int_{2m}^{\infty} dz \frac{\left(z^2+2m^2\right)\sqrt{\left(z^2-4m^2\right)}}{z^2} \tag{87}$$

where there is no summation over the two $\mu$ superscripts that appear on the right. For $\pi^{\mu\nu}(k)$ to be gauge invariant it must satisfy,

$$k_\nu \pi^{\mu\nu}(k) = 0 \tag{88}$$

The term $\pi_G^{\mu\nu}$ term is gauge invariant because $k_\nu \pi_G^{\mu\nu} = 0$. However the term $\pi_{NG}^{\mu\nu}$ is not gauge invariant because $k_\nu \pi_{NG}^{\mu\nu} \ne 0$. Therefore to get a physically valid result it is necessary to "correct" equation (85) by dropping $\pi_{NG}^{\mu\nu}$ from the solution. A similar situation exists when other sources in the literature are examined. For example consider the discussion in Section 14.2 of Greiner et al [10]. Greiner writes the solution for the vacuum polarization tensor (see equation 14.43 of [10]) as,

$$\pi^{\mu\nu}(k) = \left(g^{\mu\nu}k^2 - k^\mu k^\nu\right)\pi\left(k^2\right) + g^{\mu\nu}\pi_{sp}\left(k^2\right) \tag{89}$$

where the quantities $\pi\left(k^2\right)$ and $\pi_{sp}\left(k^2\right)$ are given in [10]. Referring to (89) it can be easily shown that the first term on the right is gauge invariant. However the second term is not gauge invariant unless $\pi_{sp}\left(k^2\right)$ equals zero. Greiner shows that this is not the case. Therefore this term must be dropped from the result in order to obtain a physically valid solution.

For another example of this problem refer to section 6-4 of Nishijima [13]. In this reference is it shown that the vacuum polarization tensor includes a non-gauge invariant term which must be removed. For other examples refer to equation 7.79 of Peskin and Schroeder [14] and Section 5.2 of Greiner and Reinhardt [15]. In all cases a direct calculation of the vacuum polarization tensor using perturbation theory produces a result which includes non-gauge invariant terms. In all cases the non-gauge invariant terms must be removed to obtain the "correct" gauge invariant result.

There are two general approaches to removing these non-gauge invariant terms. The first approach is simply to recognize that these terms, which are divergent, cannot be physically valid and drop them from the solution. This is the approach taken by Heitler [12], Nishijima [13], and Greiner et al [10]. The other approach is to come up with mathematical techniques which automatically eliminate the offending terms. This is called "regularization". There are two types of regularization. One type is called Pauli-Villars regularization [16]. In this case additional functions are introduced that have the correct behavior so that the non-gauge invariant terms are cancelled. An example of the use of Pauli-Villars regularization is given by Greiner and Reinhardt [15]. Another type of regularization is called dimensional regularization. An example of this is given by Peskin and Schroeder [14].

As we have shown QFT in the Schrödinger picture does not necessarily obey the symmetries associated with the Dirac equation. This problem was also discussed in [11] where it was shown that this is due to the fact that the quantity $\langle\Omega|\hat{H}_0|\Omega\rangle$ must be non-negative (per equation (65)). A possible solution to this problem is to redefine the vacuum state in QFT so that state vectors exist where $\langle\Omega|\hat{H}_0|\Omega\rangle$ is negative. A way to do this is presented in [4], [5], and [11]. It is shown, for example, in [4] how equivalence between hole theory and QFT is restored by properly redefining the QFT vacuum state in the case of a time independent perturbation.

## **VI. Conclusion.**

We have compared HT to QFT in the Schrödinger picture. We have shown that HT and QFT give different results and are therefore different theories. In HT the unperturbed vacuum state is not the state of minimum free field energy. It is possible to produce a state with less free field energy than the vacuum state through the interaction with a properly applied electric potential. In QFT this cannot occur because it can be shown on theoretical grounds that it is not possible to formulate a state whose free field energy is less than the free field energy of the vacuum state $|0\rangle$. In addition, due to the fact that HT is an N-electron theory, where each occupied wave function obeys the Dirac equation it will automatically obey all symmetries associated with the Dirac equation. This is not necessarily the case for QFT in the Schrödinger picture. For example, it was shown that the continuity equation does not hold in QFT in the Schrödinger picture. The result of this is the well known problem of anomalies that often occur when calculations are made. These anomalies are generally associated with divergence quantities and are eliminated when the divergences are removed by a mathematical technique called regularization.

## **Appendix**

We want to show that it is possible to find a positive electron wave function that satisfies the condition,

$$\frac{\partial J_p^{(0)}(z,t_1)}{\partial z} \neq 0 \tag{90}$$

At the initial time $t_0$ let the normalized positive electron wave function be,

$$\psi_p(z,t_0) = \frac{1}{\sqrt{2}}\left(\varphi_{+1,r}^{(0)}(z,t_0) + \varphi_{+1,s}^{(0)}(z,t_0)\right) \tag{91}$$

From (46),

$$\psi_p^{(0)}(z,t_1) = \frac{1}{\sqrt{2}}\left(\varphi_{+1,r}^{(0)}(z,t_1) + \varphi_{+1,s}^{(0)}(z,t_1)\right) \tag{92}$$

Use this in (45) to obtain,

$$J_p^{(0)}(z,t_1) = \frac{q}{2}\begin{pmatrix} \varphi_{+1,r}^{(0)\dagger}(z,t_1)\sigma_x\varphi_{+1,r}^{(0)}(z,t_1) \\ +\varphi_{+1,r}^{(0)\dagger}(z,t_1)\sigma_x\varphi_{+1,s}^{(0)}(z,t_1) \\ +\varphi_{+1,s}^{(0)\dagger}(z,t_1)\sigma_x\varphi_{+1,r}^{(0)}(z,t_1) \\ +\varphi_{+1,s}^{(0)\dagger}(z,t_1)\sigma_x\varphi_{+1,s}^{(0)}(z,t_1) \end{pmatrix} \tag{93}$$

Use $\sigma_x = \begin{pmatrix} 0 & 1 \\ 1 & 0 \end{pmatrix}$ and (4) and (5) in the above to obtain,

$$J_p^{(0)}(z,t_1) = \frac{q}{2}\begin{pmatrix} u_{+1,r}^{\dagger}\sigma_x u_{+1,r} + u_{+1,s}^{\dagger}\sigma_x u_{+1,s} \\ +u_{+1,r}^{\dagger}\sigma_x u_{+1,s} e^{i\left[(E_r-E_s)t_1-(p_r-p_s)z\right]} \\ +u_{+1,s}^{\dagger}\sigma_x u_{+1,r} e^{-i\left[(E_r-E_s)t_1-(p_r-p_s)z\right]} \end{pmatrix} \tag{94}$$

This becomes,

$$J_p^{(0)}(z,t_1) = \frac{q}{2}\begin{pmatrix} u_{+1,r}^{\dagger}\sigma_x u_{+1,r} + u_{+1,s}^{\dagger}\sigma_x u_{+1,s} \\ +2(N_{+1,r}N_{+1,s})\left(\frac{p_s}{E_s+m} + \frac{p_r}{E_r+m}\right)\cos\left((E_r-E_s)t_1-(p_r-p_s)z\right) \end{pmatrix}$$

$$\tag{95}$$

Therefore,

$$\frac{\partial J_p^{(0)}(z,t_1)}{\partial z} = q(p_r - p_s)(N_{+1,r} N_{+1,s})\left(\frac{p_s}{E_s + m} + \frac{p_r}{E_r + m}\right)\sin\left((E_r - E_s)t_1 - (p_r - p_s)z\right)$$

(96)

This is in general non-zero.